\documentclass[doublespacing]{elsart}

\usepackage{graphicx}

\usepackage{amssymb}

\begin{document}

\begin{frontmatter}



\title{A Hardware Implementation of Artificial Neural Network
Using Field Programmable Gate Arrays
}


\author{E. Won\corauthref{cor}}
\corauth[cor]{Corresponding author. 
Tel: +82 2 3290 3113; fax: +82 2 927 3292.
}
\ead{eunilwon@korea.ac.kr}

\address{
Department of Physics, Korea University, Seoul 136-713, Korea.
}

\begin{abstract}
An artificial neural network algorithm is implemented using a
field programmable gate array hardware. One hidden layer is 
used in the feed-forward neural network structure in order to
discriminate one class of patterns from the other class in real time.
With five 8-bit input patterns, six hidden nodes, and one 8-bit
output, the implemented hardware neural network makes decision
on a set of input patterns in 11 clocks and the result is
identical to what to expect from off-line computation. This
implementation may be used in level 1 hardware triggers in 
high energy physics experiments. 

\end{abstract}

\end{frontmatter}

\section{Introduction}
\label{Introduction}

 Artificial neural networks (NN) are widely used in pattern recognition problems 
in the field of particle physics experiments. Typical applications include
particle recognition in the tracking system~\cite{alice}, 
event classification problem in physics analyses~\cite{won,boos,hakl},
and hardware triggers~\cite{l3,atlas}.
For hardware triggers, realization
with standard electronics~\cite{cplear}, with dedicated NN chip~\cite{hermann},
and recently with field programmable gate arrays~\cite{atlas} (FPGA) have
been studied.
In particular, recent advances in digital technology may make it possible
to consider transferring complex level 2 NN based pattern recognition tasks to
level 1 trigger using FPGA technology. 

 In this work, a hardware
implementation of a feed-forward NN using FPGA technology is developed. 
First, training of the NN is made in offline computing environment 
in order to determine weights and thresholds of the network. And, as 
an intermediate step, 
a standalone C++ program is then written in order to discretize the NN
computation, for a bit-by-bit comparison with the response from
the hardware.  The
hardware implementation is then carried out by programming a FPGA
hardware. The performance of the implemented hardware NN and possible
application to the first level trigger in high energy
physics experiments are discussed at the end.

\section{Network Architecture}
\label{network}

\begin{figure}
\includegraphics[width=14.0cm]{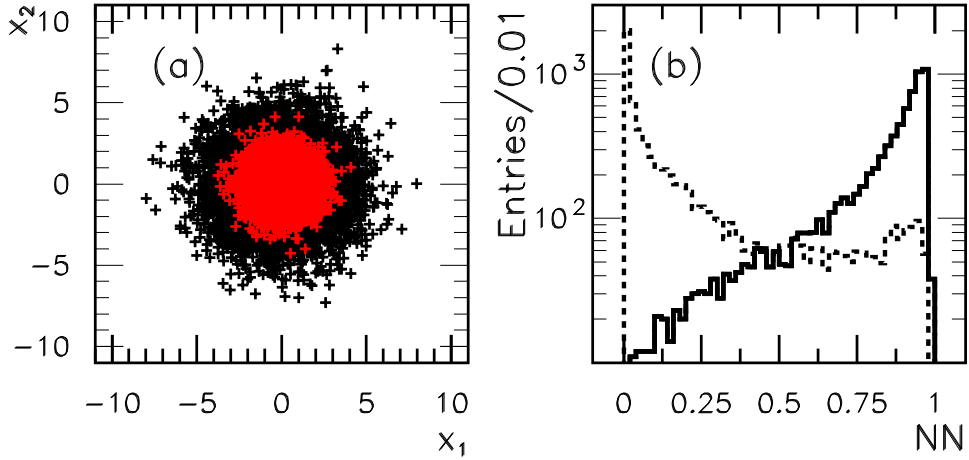}
\caption{Correlations of selected input patterns ($x_1$ and $x_2$) and
the NN output after the training are shown. In (a), the 
grey (black)-colored points are signal (background). In (b), the distribution
of the NN outputs are shown as solid and dashed histograms for the signal
and background, respectively. 
}
\label{fig:input}
\end{figure}

 A feed-forward neural network feature 
function $F_i (x_1, x_2, ..., x_N)$~\cite{nnbook} 
may be represented by the following
formula
\begin{eqnarray}
F_i (x_1, x_2, ..., x_N) = g
\left[
\frac{1}{T}
\sum_j \omega_{ij}
g
\left(
\frac{1}{T}
\sum_k \omega_{jk}
x_k + \theta_j
\right)
+ 
\theta_i
\right]
\label{eq:nn_structure}
\end{eqnarray}
where the weights $\omega_{ij}$, and thresholds $\theta_j$ are parameters
to be fitted to the entire input patterns \{$x_i$\}.
The Eq.~(\ref{eq:nn_structure}) represents one hidden-layer structure
where the first summation is over hidden nodes and the second is
over input nodes. In Eq.~(\ref{eq:nn_structure}),
$T$ is a ``temperature'' term that rescales the sum,
$N$ denotes the number of input nodes of the network, and
$g(x)$ is activation function of neurons.
The non-linear neuron activation function of the following type is 
frequently used~\cite{nnbook}
\begin{eqnarray}
g(x) = \frac{1}{1+{e}^{-2x}}
\label{eq:nn_activation}
\end{eqnarray}
in order to model the activation of the neuron. In this study,
a feed-forward neural network with 5 input patterns, first hidden later of
6 nodes and one output layer of one node is constructed (to be referred
as 5-6-1 structure from now on). In order to have a baseline NN performance,
two sets of 5-variable Gaussian random numbers are generated, one referred as
``signal'' and the other as ``background''. Training of the NN 
is carried out using JETNET~\cite{jetnet} program. 
In total, 5,000 patterns are used for the training with the cycle of 3,000.
The inverse temperature term is set to 1.0, and the back-propagation 
learning rule~\cite{bpropagation} 
is used. Even if this learning rule is rather complicated, 
it takes less then one minute with a modern personal
computer in order to train 5,000 input patterns. 
After the training, in total 36 weights and 7 threshold values
are saved in order to calculate the NN output 
on given patterns.
Figure~\ref{fig:input} (a) shows the two-dimensional distributions
of two selected input patterns for both signal and background
where there are weak correlations between patterns. Rest of the
others also have similar level of correlations. 
The trained
NN output is shown in Fig.~\ref{fig:input} (b) where the separation of 
the background from the signal is very clear. Note that, at this stage,
both input patterns and the NN output are real numbers. In particular, the
NN output is bounded in [0,1] which is the region bounded by the
activation function. 

\section{Conversions to Integer Numbers}
\label{conversion}
 In order to implement the NN algorithm in a FPGA chip, one may need to convert
the entire computational steps to integer-valued mathematical operations.
As an intermediate step toward programming a FPGA chip with a hardware
description language, a standalone C++ program that does all necessary
NN algorithm calculations in the integer-valued space
is written. This program apparently will
be a strong debugging tool when one writes the hardware
description language to program the hardware.

\begin{figure}
\includegraphics[width=14.0cm]{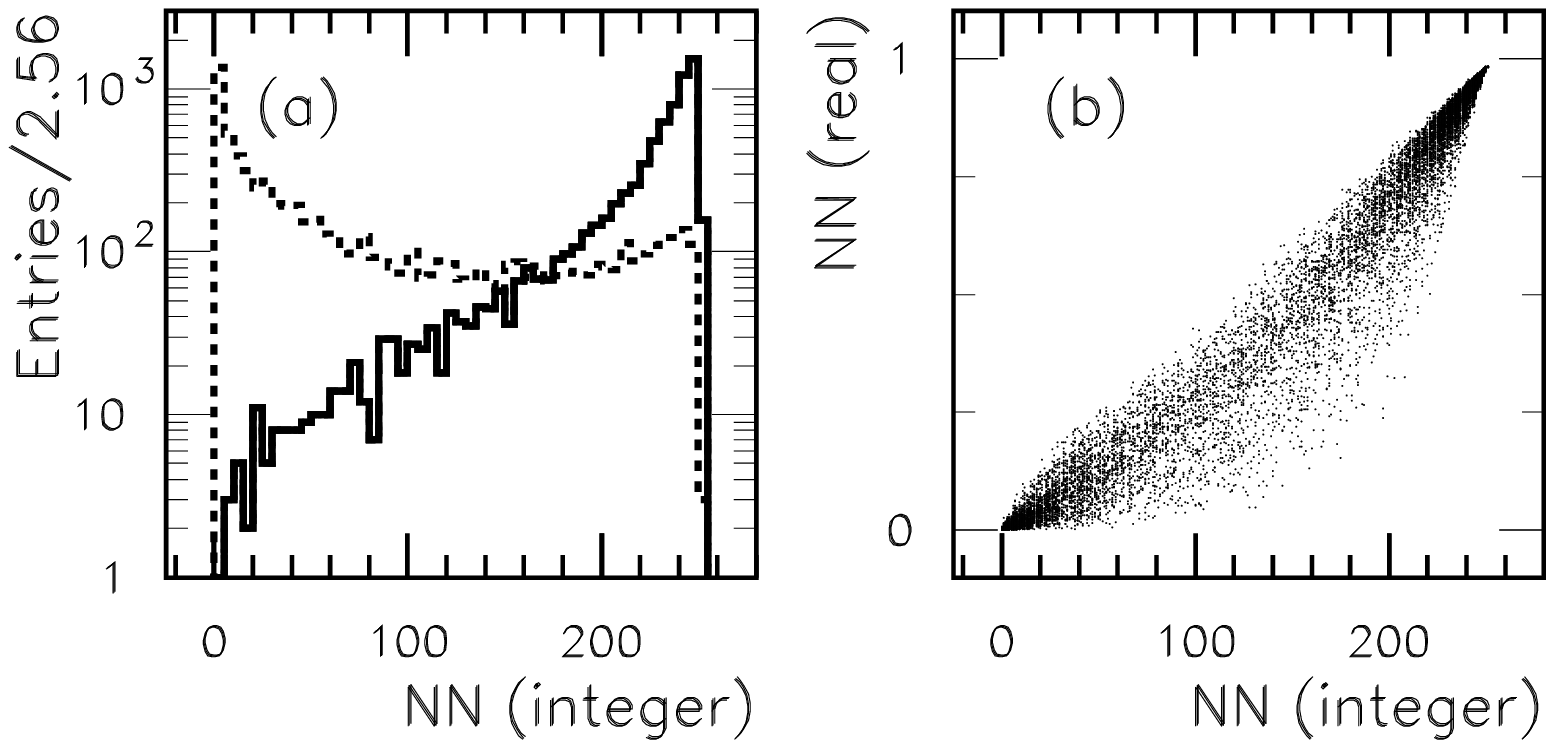}
\caption{
The output of 8 bit-wide NN is shown.
The performance of the integer-valued network with 8-bit resolution is
shown in (a) where the solid and dashed histograms are the signal
and background components, respectively. 
The correlation between the real and integer-valued (8-bit)
NN output is shown in (b). 
}
\label{fig:nn_cpp}
\end{figure}

 First, the real-valued input patterns, weights, and 
thresholds are converted into 8-bit integer
values. Here, the 8-bit is chosen so that the design of the firmware 
is appropriate for moderate performance FPGA chips available in the market.
The standalone C++ program mentioned above 
reads in the integer-valued input patterns and 
performs neural computation in purely integer-valued space with pre-stored
weights and thresholds. 
The activation function in Eq.~(\ref{eq:nn_activation})
is replaced with a 8-bit wide integer lookup table for a faster access to
the activation function at the hardware level that is implemented later. 
The NN output is also re-scaled to be bounded within
[0,2$^8$] as well. However, the internal networks storing values of
$(1/T\sum \omega_{jk}x_k + \theta_j)$ have the  bit width of 32 and therefore
little information is lost in storing values into internal networks.
By doing these conversions to integer-based calculations,
the performance of the NN output is degraded because of the 
fact that 
the conversion from to integer numbers is a round-off process and 
therefore it is natural to loose information due to such process.
Figure~\ref{fig:nn_cpp} (a) and (b) show such effects in detail.
The output of NN algorithm with integer-valued algorithm is shown in
Fig.~\ref{fig:nn_cpp} (a). The power of the discrimination can be
compared with Fig.~\ref{fig:input} (b) where the real-valued NN output 
is plotted. One can see easily that the discrimination is significantly
weaker in Fig.~\ref{fig:nn_cpp} (a). The correlation between
the real-valued versus integer-valued neural computation outcome is
shown in Fig.~\ref{fig:nn_cpp} (b). There is a strong positive correlation
between two, indicating the integer-valued version of the NN performs well,
but there are also cases where the resolution is smeared toward background-like
patterns relative to the prediction from the real-valued version.
We attribute the source of the degradation at this stage is 
purely the effect of round-off.

\begin{figure}
\includegraphics[width=14.0cm]{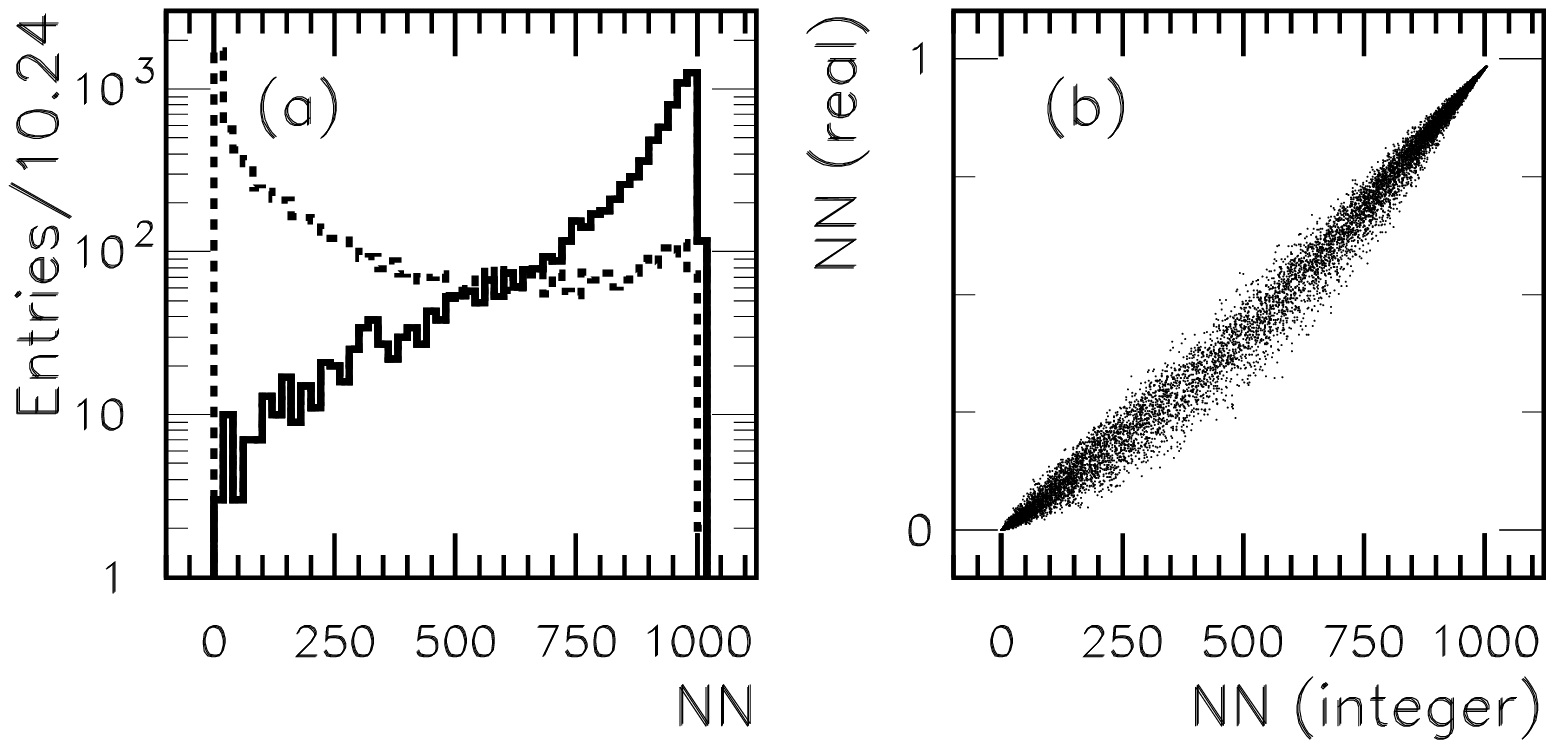}
\caption{
The output of 10 bit-wide NN is shown.
The performance of the integer-valued network with 10-bit resolution is
shown in (a) where the solid and dashed histograms are the signal
and background components, respectively. 
The correlation between the real and integer-valued (10-bit)
NN output is shown in (b). 
}
\label{fig:nn_cpp10}
\end{figure}

 In order to explicitly study the effect of the number of bits in
the discretization of the computation, the integer-valued NN with 10-bit
resolution is implemented. 
Figure.~\ref{fig:nn_cpp10} (a) shows the NN output from 
the signal and the background patterns when 10-bit resolution is
used in the computation. The discrimination power is
marginally improved compared with the 8-bit version, as expected. 
Figure.~\ref{fig:nn_cpp10} (b) 
shows the correlation between the real-valued versus 10-bit version of
the integer-valued
neural computation. Clearly,
the correlation is significantly improved 
with the 10-bit computation, supporting the argument that there
is a clear relation between the NN performance and the number of bits
used.

 A quantitative study on the performance is also carried out. Two sets
of selection criteria are applied to the real-valued version, 10-bit, and
8-bit C++ programs mentioned above. The signal-to-background
ratios are calculated
for two different cuts: NN output $>$ 0.5 and NN output $>$ 0.7 to the
real-valued NN output, and scaled cuts for integer-valued versions. Results
are summarized in Table~\ref{table:nn}. For two different selection
criteria, real-valued version
consistently gives the best signal-to-background ratio values
and are becoming worse when smaller number of bits are used in the
integer-valued neural computation. These results are consistent with the 
qualitative arguments made when figures were discussed above.

\section{Hardware Implementation}
\label{Digitization}

 A hardware implementation of the NN algorithm based on previous studies
is carried out for real online applications. The Xilinx~\cite{xilinx}
SPARTAN XC3S4000 is selected as a target FPGA. This is chosen as it has
large number of input and output pins, moderate size of random access memory 
(RAM) blocks, and configurable logic blocks. 
For the hardware implementation, the same 5-6-1 NN structure is
chosen and the C++ codes developed in the previous section is converted
into very high speed integrated circuit hardware description language
(VHDL). It has less than 800 lines of VHDL codes and 8-bit wide bus signals
are used in the network interconnections. In order to evaluate
the activation function quickly, a look-up table is implemented using
the 16-bit wide, 1,024 deep single-port block RAM that is 
available in the SPARTAN core. 
In order to reduce the resource of the given FPGA for 
future larger networks, the
symmetry in the shape of the activation function is used, resulting in
reducing the
number of the look-up tables in the design by half. 
The total number of block RAMs used in the implementation is 7 in 
5-6-1 network structure and this can easily fit in the target FPGA.  
The amount of resources spent by the neural computation logic after the
synthesis is well below 10 \% including the block RAM usage.
Several clocks are used in order to
convert integer values from/to standard logic vectors, as least for
the VHDL level simulation purpose. In total, there are 11 clocks 
required in order to produce the final NN output. This number can in 
principle be further reduced for the real time application by removing the 
conversion between integer and standard logic vectors used in the 
current VHDL for easier simulation.

\begin{figure}
\includegraphics[width=14.0cm]{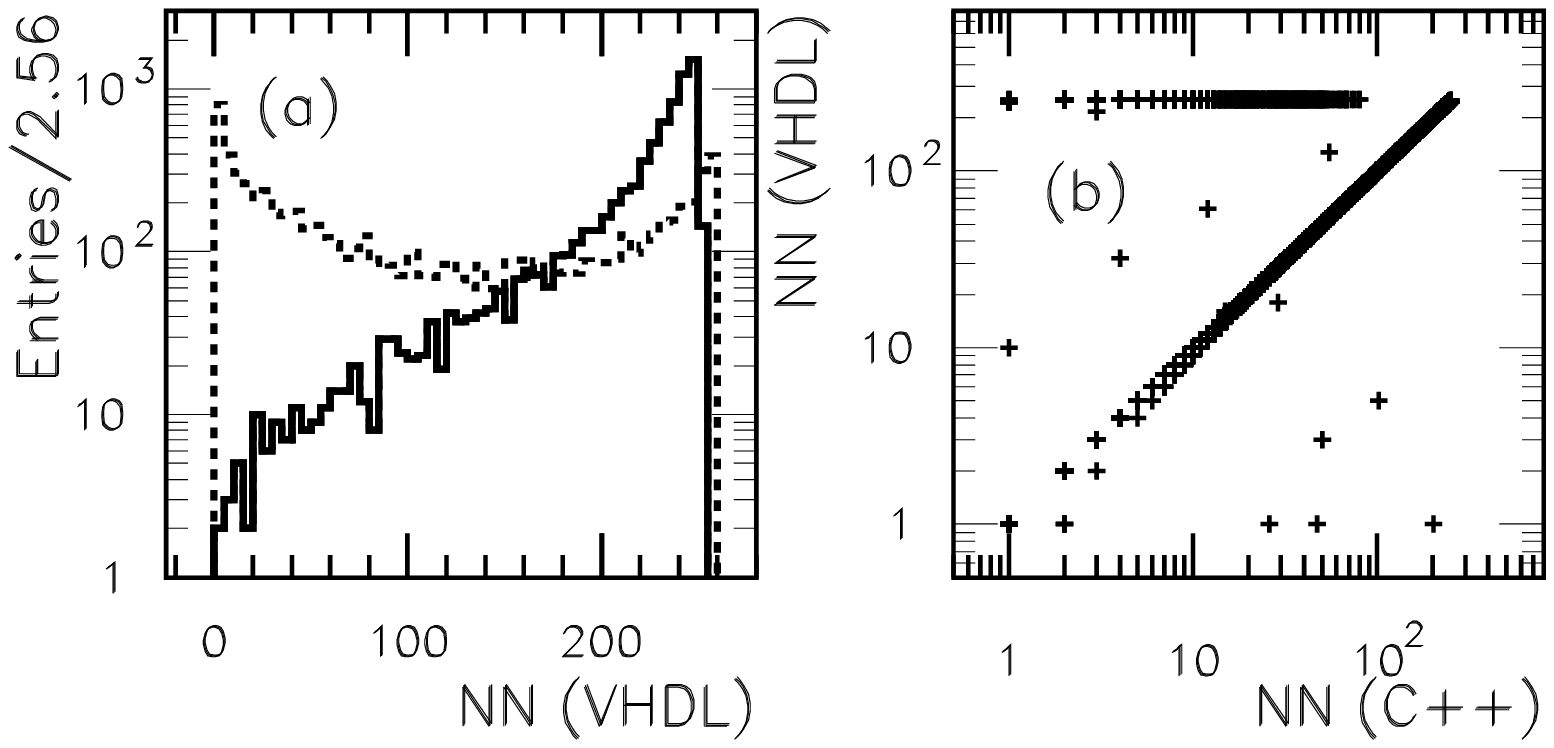}
\caption{
The output of 8 bit-wide, VHDL based NN output from the simulation is
shown in (a) where the solid and dashed histograms are the signal
and background components, respectively. 
The correlation between the VHDL and C++ based NN outputs are plotted
in (b) as log-log scale, where both are 8-bit wide computations.
}
\label{fig:nn_vhdl}
\end{figure}

 In order to check the validity of implemented 
VHDL logic, 15,000 input patterns and
NN output expectations from C++ program mentioned in the previous
section are written to a text file. The 15,000 input patterns are
fed into the VHDL logic using 
$\texttt{textio}$ package included in the standard VHDL library. 
The VHDL level simulation
for 15,000 patterns is carried out. The 
outputs are saved and their distributions are shown
in Fig.~\ref{fig:nn_vhdl} (a). The performance is degraded from the 
8-bit C++ based NN output, in particular for the left-hand side patterns
for the background. The situation is also seen from the 
Fig.~\ref{fig:nn_vhdl} (b) where the correlation between the NN output from
the VHDL based versus C++ (8-bit) based computation is compared. 
First of all, there
is a perfect correlation between two cases, indicating an excellent agreement
between two computations. Detailed study shows that,
for the diagonal patterns in Fig.~\ref{fig:nn_vhdl} (b), the spread of the 
difference is mostly zero and certainly within 3 bins out of in total 256. 
However, there are also small fraction of patterns that are accumulated
high tail of the VHDL based NN output but are populated in the
signal-like region according to 
the C++ based NN output. The detailed bit-by-bit comparison study
with our C++ program that runs over integer values 
reveals that the source of this behavior is 
out-of-range values in the integer-valued neural computation.
To be more specific, one finds that for small fraction of events,
the value of $(1/T\sum \omega_{jk}x_k +\theta_j)$
is not small enough to be 
represented with only 8 digits, and in this case, the final output
from NN becomes meaningless. Note that in the computation with 
the C++ program, such calculation is carried out with 32 digits as mentioned
earlier. This effect is also observed in the last row of Table~\ref{table:nn}
where the signal-to-noise for the VHDL based algorithm is reduced from
the value for the 8-bit version of the C++ program. This can in principle
be recovered by increasing the number of bits used in the internal networks,
but it requires the increase of the block RAMs in a multiple of the number
of hidden nodes so no further study is persued.

\section{Conclusions}
\label{Conclusions}

 A feed-forward artificial neural network 
algorithm is implemented based on the FPGA technology. A 5-6-1 neural network
is first trained with a high level language 
in a non-real time OS environment, in order to perform rather complicated
back propagation algorithm promptly. 
As an intermediate stage, a C++ program
that computes the NN output with pure integers is 
written in order to understand
the possible under-performance of the real hardware algorithm and to obtain faster
debugging of the firmware development.
It is found that the distretization process degrades the performance
of the neural network algorithm.
For the real hardware implementation, 
less than 800 lines of VHDL program is developed where
the weights and thresholds are 
distretized into 8-bit wide integers and pre-stored in the program.
The hardware level simulation shows that there is another source of the
degradation in our implementation due to the finite size of the internal
network lines during the hardware level NN computation that is unavoidable.
Nonetheless, we found that the implementation of 
a feed forward artificial neural network algorithm of three layers with
less than 10 input nodes may be easily fit in moderate FPGA chips
available in the market. An application to the level 1 hardware trigger
in a high energy physics experiment is planned.

{\bf Acknowledgments} \\ \\
E.~W.'s research was supported by grant No.
R01-2005-000-10089-0
from the Basic Research Program of the Korea Science
\& Engineering Foundation.



\newpage

\begin{table}
\begin{tabular}{c|cc}
\hline
\hline
Computational & \multicolumn{2}{c}{Signal/Background} \\
\cline{2-3}
method & NN output $>$ 0.5 & NN output $>$ 0.7 \\
\hline
JETNET (real) & 4.1 $\pm$ 0.1 & 6.3 $\pm$ 0.2   \\
C++ (10-bit) &  3.9 $\pm$ 0.1 & 5.6 $\pm$ 0.2   \\
C++ (~8-bit) &  3.4 $\pm$ 0.1 & 4.8 $\pm$ 0.1   \\
VHDL (8-bit) &  2.3 $\pm$ 0.1 & 2.8 $\pm$ 0.1   \\
\hline
\hline
\end{tabular}
\caption{\label{table:nn}
Signal-to-background ratios for 
the output of JETNET (real-valued), C++ versions (10-bit and 8-bit),
and VHDL (8-bit) version. Note that the selection cut values for 10 (8)-bit
version are 512 and 716 (128 and 179) for NN output $>$ 0.5 
and NN output $>$ 0.7, respectively.
}
\end{table}

\newpage



\begin{thebibliography}{00}
\bibitem{alice}
A. Eide, Th. Lindblad, T. Linden, and C. S. Lindsey,
Nucl. Instr. and Meth. A {\bf 368}, 855 (1996);
A. Badala, R. Barbera, G. Lo Re, A. Palmeri, G. S. Pappalardo, 
A. Pulvirenti, and F. Riggi,
Nucl. Instr. and Meth A {\bf 502}, 503 (2003).
\bibitem{won}
The D\O~collaboration, Phys. Rev. Lett. {\bf 83} 1908 (1999).
\bibitem{boos}
E. Boos, L. Dudko, and D. Smirnov,
Nucl. Instr. and Meth A {\bf 502}, 486 (2003).
\bibitem{hakl}
F. Hakl, M. Hlavacek, and R. Kalous,
Nucl. Instr. and Meth A {\bf 502}, 489 (2003).
\bibitem{l3}
D. Haas, M. Steinacher, L. Tauscher, S. Vlachos, and M. Wadhwa,
Nucl. Instr. and Meth A {\bf 420}, 101 (1999).
\bibitem{atlas}
J.-C. Prevotet, B. Denby, P. Garda, B. Granado, and C. Kiesling,
Nucl. Instr. and Meth A {\bf 502}, 511 (2003).
\bibitem{cplear}
F. R. Leimgruber, P. Pavopoulos, M. Steinacher, L. Tauscher, 
S. Vlachos, and H. Wendler,
Nucl. Instr. and Meth A {\bf 365}, 198 (1995).
\bibitem{hermann}
Hermann Kolanoski,
Nucl. Instr. and Meth A {\bf 367}, 14 (1995) and references therein.
\bibitem{nnbook}
J. Hertz, A. Krogh, and R. G. Palmer, {\it
Introduction to the Theory of Neural Computation}, Addison Wesley (1991).
\bibitem{jetnet}
C. Peterson and T. Rognvaldsson, JETNET3.0, CERN-TH.7135/94 (1994).
\bibitem{bpropagation}
D. E. Rumelhard, G. E. Hinton, and R. J. Williams,
Nature  {\bf 323}, 533 (1986).
\bibitem{xilinx}
Xilinx Corporation, San Jose, CA, USA.




\end{thebibliography}
\end{document}